\documentclass[11pt,reqno]{amsart}
\usepackage[margin=1in]{geometry}
\usepackage{amsmath,amssymb,amsthm,mathtools,bm,microtype,needspace}
\usepackage{array,booktabs,longtable}
\usepackage[hidelinks]{hyperref}
\newtheorem{theorem}{Theorem}[section]
\newtheorem{lemma}{Lemma}[section]
\newtheorem{remark}{Remark}[section]

\theoremstyle{plain}
\newcommand{\E}{\mathbb E}
\newcommand{\PP}{\mathbb P}
\newcommand{\R}{\mathbb R}
\newcommand{\F}{\mathbb F}
\newcommand{\tr}{\operatorname{tr}}
\newcommand{\diag}{\operatorname{diag}}

\newcommand{\supp}{\operatorname{supp}}
\newcommand{\eps}{\varepsilon}
\numberwithin{equation}{section}
\title{Hadamard Transforms for Dimension Reduction}

\author{Mingxuan Sun}
\address{State Key Laboratory of Mathematical Sciences, Academy of Mathematics and Systems Science, Chinese Academy of Sciences, Beijing, 100190, China; School of Mathematical Sciences, University of Chinese Academy of Sciences, Beijing, 100091, China}
\email{sunmingxuan@amss.ac.cn}

\author{Zhiqiang Xu}
\address{State Key Laboratory of Mathematical Sciences, Academy of Mathematics and Systems Science, Chinese Academy of Sciences, Beijing, 100190, China; School of Mathematical Sciences, University of Chinese Academy of Sciences, Beijing, 100091, China}
\email{xuzq@lsec.cc.ac.cn}

\begin{document}

\begin{abstract}
Using the Hadamard matrix for dimension reduction can accelerate the computation due to the fast Hadamard transform. However, the previous target dimension bounds in terms of the Gaussian width depend on the ambient dimension. In this paper, we obtain the following ambient dimension free bound using three Hadamard-diagonal blocks.  Let $H\in\mathbb R^{d\times d}$ be the normalized Walsh-Hadamard matrix, and let $D_1,D_2,D_3\in\mathbb R^{d\times d}$ be independent Rademacher diagonal matrices. We prove that for any symmetric set $T\subset \mathbb{S}^{d-1}$,
\[
\PP\left\{\sup_{\bm t\in T}\bigl|\frac dm\|P_\Omega HD_3HD_2HD_1\bm t\|_2^2-1\bigr|\le\varepsilon\right\}
\ge1-\delta
\]
whenever
\[
m\gtrsim\varepsilon^{-2}w(T)^2\log^2\frac e\delta\,\log^4\!\left(e+\frac{w(T)}\varepsilon\log\frac e\delta\right).
\]
Here $w(T)$ is the Gaussian width of $T$, and $\Omega$ is a set of $m$ distinct coordinates chosen uniformly.
This is the first ambient dimension free bound on the target dimension using Hadamard transforms, for an arbitrary symmetric set in terms of its Gaussian width.

We also prove that it is impossible to obtain an ambient dimension free bound for an arbitrary set in terms of its Gaussian width using two Hadamard-diagonal blocks.
\end{abstract}

\maketitle
\section{Introduction}

\subsection{Problem Setup}
Dimension reduction has numerous applications in data science. A standard choice for dimension reduction is a normalized Gaussian matrix. The target dimension required  for a Gaussian matrix is well understood. To state the theorem, we first introduce the Gaussian width, which provides a measure of the ``complexity'' or ``size'' of the set. For $T\subset \R^d$, the Gaussian width of $T$ is
\[
w(T):=\E \sup_{\bm{t}\in T} \langle \bm g,\bm t\rangle,
\]
where $\bm{g}\sim N(\bm 0, I_d)$ is a standard Gaussian vector.
\begin{theorem}[{\cite[Theorem~8.4]{MT20}}]\label{thm:normalmap}
Let $\eps,\delta\in (0,1)$ and $T\subset \mathbb{S}^{d-1}$. Suppose $A\in\R^{m\times d}$ is a random matrix with i.i.d. $N(0,1/m)$ entries. Then,
\[
\PP\left\{\sup_{\bm t\in T}\bigl|\|A\bm t\|_2-1\bigr|\le\eps\right\}\ge1-\delta,
\]
whenever $m\ge (2w(T)^2+6\log(2e/\delta))/\eps^2$.
\end{theorem}
However, multiplication by a dense Gaussian matrix can be computationally expensive in practice.

This motivates the use of the Hadamard transform. Let $d=2^r$, where $r$ is a positive integer. Denote the finite field of two elements by $\mathbb F_2$, and set $G:=\mathbb F_2^r$, i.e. $G:=\{0,1\}^r$.  The normalized Walsh-Hadamard matrix $H\in\R^{G\times G}$ is defined as
\[
H_{\bm{i},\bm{j}}=\sqrt{\frac 1d}(-1)^{\langle \bm i,\bm j\rangle}, \qquad \bm i,\bm j\in G,
\]
where $\langle \cdot,\cdot\rangle$ denotes the standard inner product over $\mathbb F_2$. For a Hadamard transform of order $d=2^r$, the fast Hadamard transform requires $O(d\log d)$ arithmetic operations per vector. Therefore, the Walsh-Hadamard matrix is widely used in fast randomized dimension reduction \cite{AC09,BG,MT20,PW,PW17,YPW17}. Since multiplication by a diagonal matrix is cost-free, a common choice is using a Rademacher diagonal sign matrix $D\in\R^{d\times d}$ followed by the normalized Hadamard matrix $H$, where the diagonal entries $D_{ii}\in\{1,-1\}$ are independent Rademacher random variables. We call the combination $HD$ a Hadamard-diagonal block. Compared with adding a code matrix or random sparse matrix in the embedding, using only Hadamard matrices and random diagonal matrices is easy to implement in practice and strictly preserves the computational advantage afforded by the fast Hadamard transform. Hadamard-diagonal blocks are also of theoretical interest because of their connection to random walks on the orthogonal group \cite{AilonRauhut2014,ChoromanskiRowlandWeller2017}.

Naturally, two fundamental questions arise concerning the Hadamard transform.

\noindent\textbf{Question I:}
{\itshape
Can embeddings based on Walsh-Hadamard matrices achieve the Gaussian target dimension bound $O(w(T)^2\eps^{-2})$ for an arbitrary symmetric $T\subset \mathbb{S}^{d-1}$?
\par}

\noindent\textbf{Question II:}
{\itshape
What is the minimum number of Hadamard-diagonal blocks needed to embed an arbitrary symmetric $T\subset \mathbb{S}^{d-1}$ with ambient dimension independent target dimension, in terms of its Gaussian width?
\par}

Although several prior works have focused on this direction, both questions remain open. For Hadamard-based embeddings of general sets, previously known target dimension bounds in terms of the Gaussian width depend on the ambient dimension.
We investigate both problems in this paper. For Question I, our construction with three Hadamard-diagonal blocks achieves the Gaussian target dimension bound up to logarithmic factors in $\eps$, $w(T)$, and the failure probability. For Question~II, we show that the answer is three, and the third Hadamard-diagonal block in our construction is essential.

\subsection{Related Work}
The earliest work in this direction is due to Pilanci and Wainwright \cite{PW}. Let $HD\in \R^{d\times d}$ be a Hadamard-diagonal block. Suppose $\Omega\in [d]^m$ are $m$ coordinates chosen uniformly with replacement, independent of $D$, and denote the projection matrix onto $\Omega$ by $Q_\Omega\in\R^{m\times d}$. For $\eps\in(0,1)$, $\delta=1/3$, and an arbitrary symmetric $T\subset \mathbb{S}^{d-1}$, they established a sufficient condition
on $m$ under which
\[
\PP\left\{\sup_{\bm t\in T}\bigl|\frac{d}{m}\|Q_\Omega HD\bm t\|_2-1\bigr|\le\eps\right\}\ge1-\delta.
\]

Later works \cite{ORS,XuJacques,IST} improve the bound on $m$ in \cite{PW} with new arguments. The following table summarizes these bounds. Throughout this table, we use $P_{\Omega}$ to denote choosing rows uniformly without replacement, and use $Q_{\Omega}$ to denote choosing rows uniformly with replacement.

{\footnotesize
\noindent\begin{tabular}{@{}p{1.3in}p{1.1in}l@{}}
\toprule
Reference & Matrix & Bound on the target dimension \\
\midrule
\cite[Proposition~2]{PW}
& $\sqrt{d/m}\,Q_\Omega HD$
& $m\ge C\eps^{-2}w(T)^4\log^4(ed),\quad\delta=1/3$ \\
\cite[Theorem~3.3]{ORS}
& $\sqrt{d/m}\,Q_\Omega HD$
& $m\ge C\eps^{-2}w(T)^2\log^4(ed)\log^2(2/\delta)$ \\
\cite[Proposition~3.4]{XuJacques}
& $\sqrt{d/m}\,P_\Omega HD$
& $m\ge C\eps^{-2}w(T)^2\log^4(ed)\log^2(2/\delta)$ \\
\cite[Theorems~2.4--2.5]{IST}
& $\sqrt{d/m}\,Q_\Omega HD$
& $m\ge C\eps^{-2}w(T)^2\log(2/\delta)\log(2ed/\delta)\log^2\!\bigl(C\eps^{-2}w(T)^2\log(2/\delta)\bigr)$ \\
\bottomrule
\end{tabular}\par
}

More recently, Boahen, Boedihardjo, Chiclana, and Iwen~\cite{BBCI2025} study embeddings of the form $BD_2HD_1\in\R^{m\times d}$, where $B\in\R^{m\times d}$ is a normalized four-wise independent code matrix. For an arbitrary set $T\subset \mathbb{S}^{d-1}$ and $\delta\le 1/2$, they obtained the following sufficient condition on the target dimension
\[
m\ge C\eps^{-4}\bigg(\Big(\log\frac1\delta\Big)^{3/4}+\sum_{j=0}^{\infty}2^{-j}\left(\log N\bigl(T,\|\cdot\|_2,2^{-j}\bigr)\right)^{3/4}\bigg)^2.
\]
Here $N(T,\|\cdot\|_2,\eta)$ denotes the Euclidean covering number of $T$ at radius $\eta$, and $C$ is a universal constant. This estimate has no explicit dependence on $d$, but retains an $\eps^{-4}$ dependence on the distortion and measures the complexity of $T$ through the displayed entropy sum. Note that we can control the Gaussian width by the displayed entropy sum, while the entropy sum can be arbitrarily worse than Gaussian width. Hence, it remains open whether a target dimension bound independent of the ambient dimension can be obtained in terms of Gaussian width.

\subsection{Notations}
Throughout this paper, we denote the entries of a vector $\bm{x}$ by $x_i$. For a vector $\bm{x}$, we denote the diagonal matrix whose diagonal entries are $x_i$ by $\diag(\bm{x})$. For a positive integer $d$, we write $[d]:=\{1,\dots,d\}$. For $I\subset [d]$, we define the projection matrix $P_I\in \R^{d\times d}$ as $(P_I)_{i,j}=1$ if $i=j$ and $i\in I$, $(P_I)_{i,j}=0$ otherwise. For a function $f(x,y)$ of random variables $x,y$, we use $\E_x f(x,y)$ to denote taking expectation over $x$, conditional on the other random inputs. We will explicitly specify the random variables being integrated like this in every conditional expectation. For a random variable $x$, we use $\|x\|_{L^p}$ to denote $(\E |x|^p)^{1/p}$. For a random matrix $X$, we use $\|X\|_{2,L^p}$ to denote $(\E \|X\|_2^p)^{1/p}$. For random variables $X$ and $Y$, we write $X\overset{d}{=}Y$ if they have the same distribution.
\subsection{Our Contribution}
In the following theorem, we show that using three Hadamard-diagonal blocks can provide embeddings with ambient dimension independent target dimension.

\begin{theorem}\label{thm:threeblocks}
Let $H\in\R^{d\times d}$ be the Hadamard transform, and let $D_1,D_2,D_3\in\R^{d\times d}$ be independent Rademacher diagonal matrices. Set $A_3:=HD_3HD_2HD_1$. Suppose $\Omega\subset [d]$ is a set of $m$ distinct coordinates chosen uniformly from $[d]$, independently of $D_1,D_2,D_3$. Then for any symmetric set $T\subset \mathbb{S}^{d-1}$, $\eps,\delta\in(0,1)$, we have
\begin{equation}\label{eq:normbound}
\PP\left\{\sup_{\bm t\in T}\bigl|\frac dm\|P_\Omega A_3\bm t\|_2^2-1\bigr|\le\eps\right\}\ge1-\delta
\end{equation}
whenever
\begin{equation}\label{eq:samplesize}
 m\ge C_1\eps^{-2}w(T)^2\log^2\frac e\delta\log^4\!\left(e+\frac{w(T)}\eps\log\frac e\delta\right).
\end{equation}
Here $C_1$ is a universal constant.
\end{theorem}

The target dimension bound in Theorem~\ref{thm:threeblocks} matches the bound in Theorem~\ref{thm:normalmap} up to logarithmic factors in $\eps, w(T),\delta$. Hence, we prove that Hadamard transforms can provide comparable embeddings to the Gaussian matrix.

In the next theorem, we construct a counterexample for two Hadamard-diagonal blocks.

\begin{theorem}\label{thm:twoblocks}
Suppose $D_1,D_2\in\R^{d\times d}$ are independent Rademacher diagonal matrices. For every fixed integer $m\ge1$, there exists a symmetric set $T_{m,r}\subset \mathbb{S}^{d-1}$ defined for all
sufficiently large $r$ with $w(T_{m,r})<21$ and
\begin{equation}\label{eq:blockfailure}
 \lim_{r\to\infty} \sup_{\Omega\subset G, |\Omega|=m}\PP\left\{ \sup_{\bm t\in T_{m,r}}\bigl|\frac{d}{m}\|P_\Omega HD_2HD_1\bm t\|_2^2-1\bigr|\le 1\right\}=0.
\end{equation}
\end{theorem}
Since the $m$ in Theorem~\ref{thm:twoblocks} is arbitrary, Theorem~\ref{thm:twoblocks} implies that for embeddings constructed from two Hadamard-diagonal blocks, any target dimension bound in terms of its Gaussian width must depend on the ambient dimension. Therefore, Theorem~\ref{thm:twoblocks} shows the third Hadamard-diagonal block in Theorem~\ref{thm:threeblocks} plays an important role in the embedding. Combining Theorem~\ref{thm:twoblocks} and Theorem~\ref{thm:threeblocks}, we prove the answer to Question~II is three.

\begin{remark}
In practical applications, $T$ is often a finite set. If $T\subset \mathbb{S}^{d-1}$ is a set of $N$ points, then $w(T)\le \sqrt{2\log N}$. Such target dimension bounds in terms of $\log N$ are covered by the literature. However, data encountered in practice often has additional algebraic or geometric structures. Gaussian width can capture such structures well, leading to a much smaller quantity than the rough $\sqrt{2\log N}$ estimate. Hence, a target dimension bound in terms of the Gaussian width overcomes the limitation that previous results missed the intrinsic structure of $T$. Therefore, Theorem~\ref{thm:twoblocks} and Theorem~\ref{thm:threeblocks} not only have theoretical significance but also provide practical guidance for algorithm design.
\end{remark}

\section{Proof of Theorem~\ref{thm:threeblocks}}\label{sec:mainproof}

We prove Theorem~\ref{thm:threeblocks} by separating the argument according to the ambient dimension. If the ambient dimension $d$ is small, we can use the following ambient dimension dependent result directly.
\begin{lemma}[{\cite[Proposition~3.4]{XuJacques}}]\label{lem:singleblock}
Let $D$ be a Rademacher diagonal matrix. Suppose $\Omega$ is a set of $m$ coordinates chosen uniformly, independently of $D$. Assume $m\ge C_2\eps^{-2}w(T)^2\log^4(ed)\log^2(e/\delta)$. Then for every symmetric $T\subset \mathbb{S}^{d-1}$ and $\eps,\delta\in(0,1)$, we have
\begin{equation}\label{eq:blockbound}
\PP\left\{\sup_{\bm t\in T}\left|\frac dm\|P_\Omega HD\bm t\|_2^2-1\right|\le\eps\right\}\ge1-\delta.
\end{equation}
\end{lemma}

If the ambient dimension $d$ is large, we develop the following estimate.

\begin{lemma}\label{lem:largespace}
Suppose $\Omega$ is a set of $m$ coordinates chosen uniformly with $d\ge(C_3m)^{40}$, independently of $D_1,D_2,D_3$. Assume $m\ge C_4\eps^{-2} \left(w(T)^2+\log(e/\delta)\right) \log^3(em)$. Then for every symmetric $T\subset \mathbb{S}^{d-1}$ and $\eps,\delta\in(0,1)$, we have
\[
 \PP\left\{\sup_{\bm t\in T}\bigl|\frac dm\|P_\Omega A_3\bm t\|_2^2-1\bigr|\le\eps\right\}\ge1-\delta.
\]
Here $C_4,C_3$ are universal constants.
\end{lemma}

The proof of Lemma~\ref{lem:largespace} is deferred to the next subsection. We now prove Theorem~\ref{thm:threeblocks}.

\begin{proof}[Proof of Theorem~\ref{thm:threeblocks}]
Observe that for every $C_4>0$, there exists $C_1$ depending only on $C_4$, such that $m\ge C_1w\log^4w$ implies $m\ge C_4w\log^4m$ for all $w\ge e$.
If $d\le(C_3m)^{40}$, then $\log(d)\lesssim\log(em)$. Under this assumption, combining Lemma~\ref{lem:singleblock} with the observation above gives, with the hypotheses in Theorem~\ref{thm:threeblocks},
\begin{equation}\label{eq:directbound}
\PP\left\{\sup_{\bm t\in T}\bigl|\frac dm\|P_\Omega HD_3\bm t\|_2^2-1\bigr|>\eps\right\}\le\delta.
\end{equation}
Since $HD_2HD_1$ is orthogonal, $w(HD_2HD_1T)=w(T)$. Since \eqref{eq:directbound} is independent of $D_1$ and $D_2$, applying \eqref{eq:directbound} to each $HD_2HD_1T$ proves Theorem~\ref{thm:threeblocks} under $d\le(C_3m)^{40}$.

If $d>(C_3m)^{40}$, then Theorem~\ref{thm:threeblocks} follows from Lemma~\ref{lem:largespace}. Therefore, we prove Theorem~\ref{thm:threeblocks} for both parameter ranges and the proof is completed.
\end{proof}

\subsection{Proof of Lemma~\ref{lem:largespace}}\label{subsec:largeproof}

We first establish several auxiliary lemmas, and prove Lemma~\ref{lem:largespace} at the end of this subsection.
The following lemma is a standard property of the Gaussian width.

\begin{lemma}[{\cite[Lemma~2.6]{BBG}}]\label{lem:normalspread}
Let $\bm g\in\R^G$ be a standard Gaussian vector. For every $u>0$ and symmetric $T\subset \mathbb{S}^{d-1}$,
\begin{equation}\label{eq:widthtail}
 \PP\left\{\left|\sup_{\bm t\in T}|\langle\bm g,\bm t\rangle|-w(T)\right|>u\right\}
 \le2e^{-u^2/2}.
\end{equation}
\end{lemma}

The following lemma bounds the fluctuations of a function of a Gaussian vector through its derivative.

\begin{lemma}[{\cite[Theorem~2.2]{Pisier}}]
\label{lem:normalchange}
Let $\bm g_1,\bm g_2\in\R^G$ be independent standard Gaussian vectors. Suppose $f:\R^G\to\R^G$ is globally Lipschitz and $\Phi:\R^G\to\R$ is continuous and convex,  with at most polynomial growth. Then
\begin{equation}\label{eq:normalchange}
\E\Phi\!\left(f(\bm g_1)-\E f(\bm g_1)\right)
\le\E\Phi\!\left(\frac\pi2\,J_f(\bm g_1)\bm g_2\right),
\end{equation}
where $J_f$ is the Jacobian matrix, defined almost
everywhere.
\end{lemma}

We next establish moment bounds on the operator norm of an associated matrix in our proof.

\begin{lemma}\label{lem:largeentries}
Set $k:=\lceil m+\log d\rceil$. We call a matrix $K\in\R^{G\times G}$ a Walsh-circulant if $K_{\bm x,\bm y}=K_{\bm0,\bm x+\bm y}$. Let $\Sigma_0\in\R^{G\times G}$ be a positive semidefinite Walsh-circulant matrix with $(\Sigma_0)_{\bm i,\bm i}=1$, $\max_{\bm i\in G}\sum_{\bm j\in G}|(\Sigma_0)_{\bm i,\bm j}|\le m$, and $d\ge(C_3m)^{40}$. Let $D$ be a Rademacher diagonal matrix, and set $\Sigma_1:=HD\Sigma_0DH$. Given $\Sigma_1$, suppose $\bm z\sim N(\bm{0},\Sigma_1)$ is a Gaussian vector. Define
\[
I:=\{\bm i\in G:|{z}_{\bm i}|>C_2\sqrt{\log(em)}\}.
\]
Then we have
\begin{equation}\label{eq:entrymoment}
 \|P_I\Sigma_1P_I\|_{2,L^k}\le4.
\end{equation}
Here $C_2$ is a universal constant.
\end{lemma}

The proof of Lemma~\ref{lem:largeentries} is deferred to the next subsection. We next prove a moment bound using Lemma~\ref{lem:largeentries}.

\begin{lemma}\label{lem:scalarmoment}
Suppose $m\in[d]$ satisfies $d\ge(C_3m)^{40}$ and $1\le s\le m$. Let $\bm z\in\R^G$ be a unit vector with $|\supp\bm z|\le m$. Then for every symmetric $T\subset \mathbb{S}^{d-1}$, \begin{equation}\label{eq:sparsemoment}
 \left\|\sup_{\bm t\in T}|\langle\bm z,\sqrt d\,A_3\bm t\rangle|\right\|_{L^s}
 \lesssim\sqrt{\log(em)}\bigl(w(T)+\sqrt s\bigr).
\end{equation}
\end{lemma}

\begin{proof}
Since $T$ is symmetric, $\E\sup_{\bm t\in T}|\langle\bm g,\bm t\rangle|=w(T)$.
Integrating the tail bound in Lemma~\ref{lem:normalspread} gives
\[
 \E\Big(\sup_{\bm t\in T}|\langle\bm g,\bm t\rangle|-w(T)\Big)_+^s
 \le2s\int_0^\infty u^{s-1}e^{-u^2/2}\,du
 =s2^{s/2}\Gamma(s/2).
\]
Since $(s2^{s/2}\Gamma(s/2))^{1/s}\lesssim\sqrt s$, we have
\begin{equation}\label{eq:widthmoment}
 \left\|\sup_{\bm t\in T}|\langle\bm g,\bm t\rangle|\right\|_{L^s}
 \lesssim w(T)+\sqrt s.
\end{equation}

Set $C_{\bm z}:=\sqrt d HD_2H\diag(H\bm z)H$. Let $\bm g_1,\bm g_2\in\R^G$ be independent standard Gaussian vectors, independent of $D_1,D_2,D_3$. Write $D_3:=\diag(\bm{\xi}^{(3)})$. Then
\[\langle\bm z,\sqrt d\,A_3\bm t\rangle
 =\left\langle\bm\xi^{(3)},\sqrt d\,\diag(H\bm z)HD_2HD_1\bm t\right\rangle,\qquad
 \E_{\bm{g}_1}\diag(\bm\xi^{(3)})|\bm{g}_1|=\sqrt{2/\pi}\,\bm\xi^{(3)}.
\]
Applying Jensen's inequality yields
\[
 \begin{split}
 \E_{D_3}\sup_{\bm t\in T}|\langle\bm z,\sqrt{d}A_3\bm t\rangle|^s &= (\pi/2)^{s/2}\E_{D_3}\sup_{\bm t\in T}\left|\left\langle\E_{\bm{g}_1}\diag(\bm\xi^{(3)})|\bm{g}_1|, \sqrt d\,\diag(H\bm z)HD_2HD_1\bm t\right\rangle\right|^s \\
 &\le(\pi/2)^{s/2}\E_{D_3,\bm g_1}\sup_{\bm t\in T}
 \left|\left\langle\diag(\bm\xi^{(3)})|\bm{g}_1|,\sqrt d\,\diag(H\bm z)HD_2HD_1\bm t\right\rangle\right|^s\\
 &=(\pi/2)^{s/2}\E_{\bm g_1}\sup_{\bm t\in T}|\langle D_1C_{\bm z}\bm g_1,\bm t\rangle|^s.
 \end{split}
\]
Then taking expectation with respect to $D_1,D_2$ gives
\begin{equation}\label{eq:normalreplace}
\left\|\sup_{\bm t\in T}|\langle\bm z,\sqrt d\,A_3\bm t\rangle|\right\|_{L^s}
\le\sqrt{\pi/2}\left\|\sup_{\bm t\in T}|\langle D_1C_{\bm z}\bm g_1,\bm t\rangle|\right\|_{L^s}.
\end{equation}

For $x\in\R$, define $f(x):=\operatorname{sign}(x)(|x|-C_2\sqrt{\log(em)})_+$, and for a vector $\bm{x}$, define $f(\bm{x})$ by applying $f$ coordinatewise. Set $\bm v:=C_{\bm{z}}\bm{g}_1$ and $\bm u:=\bm v-f(\bm v)$. We claim that
\begin{equation}\label{eq:boundedmoment}
 \left\|\sup_{\bm t\in T}|\langle D_1\bm u,\bm t\rangle|\right\|_{L^s}
\lesssim\sqrt{\log(em)}\bigl(w(T)+\sqrt s\bigr),
\end{equation}
\begin{equation}\label{eq:excessmoment}
 \left\|\sup_{\bm t\in T}|\langle D_1f(\bm v),\bm t\rangle|\right\|_{L^s}
 \lesssim w(T)+\sqrt s.
\end{equation}
Then substituting \eqref{eq:boundedmoment} and \eqref{eq:excessmoment} into \eqref{eq:normalreplace} proves \eqref{eq:sparsemoment}.

It remains to prove the two claims. We first prove \eqref{eq:boundedmoment}. Given $\bm u$, let $\bm h\sim N(\bm{0},C_2^2\log(em)I_d-\diag(\bm u)^2)$ be a random Gaussian vector independent of $\bm g_2$. Then $\diag(\bm u)\bm g_2+\bm h\overset{\mathrm d}=C_2\sqrt{\log(em)}\bm{g}_2$. Applying Jensen's inequality gives
\[
 \begin{split}
 &\left\|\sup_{\bm t\in T}|\langle D_1\bm u,\bm t\rangle|\right\|_{L^s(D_1)} \le\sqrt{\pi/2} \left\|\sup_{\bm t\in T}|\langle\diag(\bm u)\bm g_2,\bm t\rangle|\right\|_{L^s(\bm g_2)}\\
 =& \sqrt{\pi/2} \left\|\sup_{\bm t\in T}|\langle\diag(\bm u)\bm g_2+\E_{\bm{h}}\bm{h},\bm t\rangle|\right\|_{L^s(\bm g_2)}
 \le\sqrt{\pi/2}\left\|\sup_{\bm t\in T}|\langle\diag(\bm u)\bm g_2+\bm h,\bm t\rangle|\right\|_{L^s(\bm g_2,\bm h)}\\
 \lesssim &\sqrt{\log(em)} \left\|\sup_{\bm t\in T}|\langle\bm g_2,\bm t\rangle|\right\|_{L^s(\bm g_2)}\lesssim\sqrt{\log(em)}\bigl(w(T)+\sqrt s\bigr).
 \end{split}
\]
Then taking expectation with respect to $D_2,\bm g_1$ proves \eqref{eq:boundedmoment}.

We next prove \eqref{eq:excessmoment}. Set $I:=\supp f(\bm{v})$.
Note that
\[
(\sqrt d H\diag(H\bm z)H)_{\bm i,\bm j} =d^{-1}\sum_{\bm x,\bm y\in G}(-1)^{\bm y\cdot(\bm i+\bm j+\bm x)} z_{\bm x}= z_{\bm i+\bm j}.
\]
Set $\Sigma_0:=d H\diag(H\bm z)H(H\diag(H\bm z)H)^{\mathsf T}$. Then $(\Sigma_0)_{\bm i,\bm j} =\sum_{\bm x\in G} z_{\bm x} z_{\bm x+\bm i+\bm j}$. Hence, $\Sigma_0$ is Walsh-circulant. Moreover, $(\Sigma_0)_{\bm i,\bm i}=\|\bm z\|_2^2=1$ and $\sum_{\bm j\in G}|(\Sigma_0)_{\bm i,\bm j}|\le\sum_{\bm j,\bm x\in G}| z_{\bm i+\bm x} z_{\bm j+\bm x}|\le|\supp \bm z|\,\|\bm z\|_2^2\le m$. For fixed $D_2$, the covariance of $\bm v$ is $C_{\bm z}C_{\bm z}^{\mathsf T}=HD_2\Sigma_0D_2H$.
Since $d\ge(C_3m)^{40}$, applying Lemma~\ref{lem:largeentries} yields
\begin{equation}\label{eq:rowmoment}
\|P_IC_{\bm z}\|^2_{2,L^{2m}}=\|P_IC_{\bm z}C_{\bm z}^{\mathsf T}P_I\|_{2,L^m}
\le 4.
\end{equation}

Fix $D_1,D_2$, and set $f_1(\bm x):=D_1f(C_{\bm z}\bm x)$, $\Phi(\bm x):=\sup_{\bm t\in T}|\langle\bm x,\bm t\rangle|^s$. Then $f_1$ is globally Lipschitz. $\Phi$ is continuous and convex, with polynomial growth. Moreover, we have $J_{f_1}(\bm g_1)=D_1P_IC_{\bm z}$ almost surely. Since $f_1$ is an odd function, $\E_{\bm g_1}f_1(\bm g_1)=\bm0$. Applying Lemma~\ref{lem:normalchange} yields
\begin{equation}\label{eq:excesschange}
\left\|\sup_{\bm t\in T}|\langle D_1f(\bm v),\bm t\rangle|\right\|_{L^s(\bm g_1)}\le\frac\pi2
\left\|\sup_{\bm t\in T}|\langle D_1P_IC_{\bm z}\bm g_2,\bm t\rangle|\right\|_{L^s(\bm g_1,\bm g_2)}.
\end{equation}
For fixed $\bm g_1$, let $\bm h$ be a centered Gaussian vector with covariance $\|P_IC_{\bm z}\|_2^2I_d-D_1P_IC_{\bm z}C_{\bm z}^{\mathsf T}P_ID_1$, independent of $\bm g_2$. Then $D_1P_IC_{\bm z}\bm g_2+\bm h\overset{\mathrm d}=\|P_IC_{\bm z}\|_2\bm g$. Applying Jensen's inequality yields
\begin{equation}\label{eq:normalmoment}
\begin{split}
&\left\|\sup_{\bm t\in T}|\langle D_1P_IC_{\bm z}\bm g_2,\bm t\rangle|\right\|_{L^s(\bm g_2)} = \left\|\sup_{\bm t\in T}|\langle D_1P_IC_{\bm z}\bm g_2+\E_{\bm{h}}\bm{h},\bm t\rangle|\right\|_{L^s(\bm g_2)} \\
\le &\left\|\sup_{\bm t\in T}|\langle D_1P_IC_{\bm z}\bm g_2+\bm h,\bm t\rangle|\right\|_{L^s(\bm g_2,\bm h)}=\|P_IC_{\bm z}\|_2\left\|\sup_{\bm t\in T}|\langle\bm g,\bm t\rangle|\right\|_{L^s(\bm g)} \\
\lesssim &\|P_IC_{\bm z}\|_2\bigl(w(T)+\sqrt s\bigr).
\end{split}
\end{equation}
Since $s\le m$, combining \eqref{eq:normalmoment} and \eqref{eq:rowmoment} with \eqref{eq:excesschange} and taking expectation with respect to $D_1$ and $D_2$ yields \eqref{eq:excessmoment}.
This completes the proof.
\end{proof}

The following lemma bounds the cardinality of a $1/2$-net of $\mathbb{S}^{n-1}$.

\begin{lemma}[{\cite[Corollary~4.2.13]{HDP}}]\label{lem:spherenet}
For every integer $n\ge1$, there is a set
$\mathcal N\subset \mathbb{S}^{n-1}$ with $|\mathcal N|\le5^n$ such that for every $\bm x\in \mathbb{S}^{n-1}$,
\[
 \min_{\bm z\in\mathcal N}\|\bm x-\bm z\|_2\le\frac12.
\]
\end{lemma}

In the next lemma we bound the moment of large coordinates using Lemma~\ref{lem:scalarmoment} and Lemma~\ref{lem:spherenet}.

\begin{lemma}\label{lem:excesssquares}
Set $s:=\log(e/\delta)$ and define $f(x):=(x^2-C_5^2\log^2(em))_+$. Then under the hypotheses in Lemma~\ref{lem:largespace}, for every fixed $m$-element set $\Omega\subset G$,
\begin{equation}\label{eq:sampleexcess}
 \left\|\sup_{\bm t\in T}\frac1m \sum_{\bm i\in\Omega}f\bigl((\sqrt d\,A_3\bm t)_{\bm i}\bigr)\right\|_{L^s(D_1,D_2,D_3)}
 \lesssim\frac{( w(T)^2+s)\log^2(em)}m.
\end{equation}
Here $C_5$ is a universal constant.
\end{lemma}

\begin{proof}
Fix any $\Omega\subset G$ with $|\Omega|=m$. For each $I\subset\Omega$ with $|I|=\lceil w(T)^2+s\rceil$, Lemma~\ref{lem:spherenet} gives a $1/2$-net $\mathcal N_I$ of $\{\bm z\in\R^G:\operatorname{supp}\bm z\subset I,\ \|\bm z\|_2=1\}$ with $|\mathcal N_I|\le5^{\lceil w(T)^2+s\rceil}$. Then for any $\bm{x}\in\R^G$ that is supported on $I$, we have
\begin{equation}\label{eq:netbound}
 \|\bm x\|_2\le 2\max_{\bm z\in\mathcal N_I}|\langle\bm x,\bm z\rangle|.
\end{equation}
Set $\mathcal N:=\bigcup_{I\subset\Omega,|I|=\lceil w(T)^2+s\rceil}\mathcal N_I$. Define the event $\mathcal E$ as
\[
 \mathcal E:=\left\{\max_{\bm z\in\mathcal N}\sup_{\bm t\in T}|\langle\bm z,\sqrt d\,A_3\bm t\rangle|\le\frac{C_5}{2}\sqrt{ w(T)^2+s}\log(em)\right\}.
\]
On the event $\mathcal E$, \eqref{eq:netbound} gives
\begin{equation}\label{eq:energybound}
 \sup_{\bm t\in T}
 \max_{\substack{I\subset\Omega\\|I|=\lceil w(T)^2+s\rceil}}d\|P_I A_3\bm t\|_2^2
 \le C_5^2( w(T)^2+s)\log^2(em).
\end{equation}
Consequently, for every $\bm t\in T$, on the event $\mathcal E$, we have
\[
\left|\left\{\bm i\in\Omega: f((\sqrt d\,A_3\bm t)_{\bm i})>0\right\}\right| \le w(T)^2+s.
\]
Therefore, on the event $\mathcal E$, we have
\begin{equation}\label{eq:goodexcess}
\sup_{\bm t\in T}\sum_{\bm i\in\Omega}f\bigl((\sqrt d\,A_3\bm t)_{\bm i}\bigr)
\le\sup_{\bm t\in T}\max_{\substack{I\subset\Omega\\|I|=\lceil{w(T)^2+s}\rceil}}d\|P_I A_3\bm t\|_2^21_{\mathcal E}\le C_5^2( w(T)^2+s)\log^2(em).
\end{equation}

We next bound the probability of the event $\mathcal E$. Set $q:=( w(T)^2+s)\log(em)$. Note that $|\mathcal N|\le5^{\lceil w(T)^2+s\rceil}\binom m{\lceil w(T)^2+s\rceil}\le e^{3q}$. Note that the condition on $m$ in Lemma~\ref{lem:largespace} with a large enough $C_4$ gives $4q\le m$.
Since every $\bm z\in\mathcal N$ has at most $m$ nonzero coordinates, applying Lemma~\ref{lem:scalarmoment} and choosing $C_5$ sufficiently large yield
\begin{equation}\label{eq:cutoffchoice}
 \begin{split}
 &\left\|\max_{\bm z\in\mathcal N}\sup_{\bm t\in T}|\langle\bm z,\sqrt d\,A_3\bm t\rangle|\right\|_{L^{4q}(A_3)}\le |\mathcal N|^{1/(4q)}\max_{\bm z\in\mathcal N}\left\|\sup_{\bm t\in T}|\langle\bm z,\sqrt d\,A_3\bm t\rangle|\right\|_{L^{4q}(A_3)}\\
 \lesssim &e^{3/4}\sqrt{\log(em)}\bigl(w(T)+\sqrt{4q}\bigr)
 \le \frac{C_5}{2e}\sqrt{ w(T)^2+s}\log(em),
 \end{split}
\end{equation}
where the last inequality follows from $q:=( w(T)^2+s)\log(em)$.
 Applying Markov's inequality with \eqref{eq:cutoffchoice} yields
\begin{equation}\label{eq:eventtail}
 \PP_{D_1,D_2,D_3}(\mathcal E^c)
 \le\left(\frac{C_5\sqrt{ w(T)^2+s}\log(em)/(2e)}{C_5\sqrt{ w(T)^2+s}\log(em)/2}\right)^{4q}
 =e^{-4q}.
\end{equation}

Note that, without restricting to $\mathcal E$, \eqref{eq:netbound} yields
\begin{equation}\label{eq:roughbound}
\sup_{\bm t\in T}\sum_{\bm i\in\Omega}f\bigl((\sqrt d\,A_3\bm t)_{\bm i}\bigr)
\le m\sup_{\bm t\in T}\max_{\bm i\in\Omega}|(\sqrt d\,A_3\bm t)_{\bm i}|^2
\le 4m\left(\max_{\bm z\in\mathcal N}\sup_{\bm t\in T}|\langle\bm z,\sqrt d\,A_3\bm t\rangle|\right)^2.
\end{equation}
Since $s<q$, combining \eqref{eq:roughbound} with \eqref{eq:cutoffchoice} and \eqref{eq:eventtail} gives
\begin{equation}\label{eq:badexcess}
 \begin{split}
 \|1_{\mathcal E^c}\,\cdot\sup_{\bm t\in T}\sum_{\bm i\in\Omega}f\bigl((\sqrt d\,A_3\bm t)_{\bm i}\bigr)\|_{L^s(A_3)}
 &\le 4m \left\|\max_{\bm z\in\mathcal N}\sup_{\bm t\in T}|\langle\bm z,\sqrt d\,A_3\bm t\rangle|\right\|_{L^{4s}(A_3)}^2\PP(\mathcal E^c)^{1/(2s)}\\
 &\lesssim m q\log(em)e^{-2q/s}
 \le q\log(em),
 \end{split}
\end{equation}
where the last inequality uses $q/s\ge\log(em)$. Recall that $q=( w(T)^2+s)\log(em)$. Combining \eqref{eq:goodexcess} and \eqref{eq:badexcess} proves \eqref{eq:sampleexcess}.
\end{proof}

We next establish the moment bound of small coordinates. The following lemma is a standard contraction inequality.

\begin{lemma}[{\cite[Lemma~8]{Adamczak}}]\label{lem:signbound}
Let $V\subset\R^m$ be nonempty and bounded, and let
$\bm\xi\in\{-1,1\}^m$ have independent Rademacher coordinates.
Let $L\ge0$ and let $\varphi:\R\to\R$ satisfy $\varphi(0)=0$ and
$|\varphi(x)-\varphi(y)|\le L|x-y|$ for all $x,y\in\R$.
For every $s\ge1$,
\begin{equation}\label{eq:signbound}
\left\|\sup_{\bm v\in V}\Big|\sum_{j=1}^m\xi_j\varphi( v_j)\Big|\right\|_{L^s}
\le2L\left\|\sup_{\bm v\in V}\Big|\sum_{j=1}^m\xi_j v_j\Big|\right\|_{L^s}.
\end{equation}
\end{lemma}

\begin{lemma}\label{lem:clippedsample}
Let $1\le s\le m$ and define $h(x):=\min(x^2,C_5^2\log^2(em))$. Then under the hypotheses in Lemma~\ref{lem:largespace}, we have
\begin{equation}\label{eq:clippedmoment}
\left\|\sup_{\bm t\in T}\Big|\frac1m\sum_{\bm i\in\Omega}h\bigl((\sqrt d\,A_3\bm t)_{\bm i}\bigr)-\frac1d\sum_{\bm i\in G}h\bigl((\sqrt d\,A_3\bm t)_{\bm i}\bigr)\Big|\right\|_{L^s}
\lesssim\sqrt{\frac{w(T)^2+s}{m}}(\log(em))^{3/2}.
\end{equation}
\end{lemma}

\begin{proof}
Given $\Omega$, choose $\Omega'$ uniformly among the $m$-element subsets of $G\setminus\Omega$. Since $\PP\{\bm i\in\Omega'\}=m/(d-m)$ for $\bm i\notin\Omega$ and fixed $\Omega$, we have, for fixed $D_1$, $D_2$, $D_3$, and $\bm t\in T$,
\begin{equation}\label{eq:sampleaverage}
\begin{split}
&\frac1m\sum_{\bm i\in\Omega}h\bigl((\sqrt d\,A_3\bm t)_{\bm i}\bigr)-\frac1d\sum_{\bm i\in G}h\bigl((\sqrt d\,A_3\bm t)_{\bm i}\bigr)\\
=&\frac{d-m}{md}\left(\sum_{\bm i\in\Omega}h\bigl((\sqrt d\,A_3\bm t)_{\bm i}\bigr)-\E_{\Omega'}\sum_{\bm i\in\Omega'}h\bigl((\sqrt d\,A_3\bm t)_{\bm i}\bigr)\right).
\end{split}
\end{equation}
Write $(\Omega,\Omega')$ as $(\bm i_1,\ldots,\bm i_m,\bm j_1,\ldots,\bm j_m)$, where the elements are ordered uniformly at random. Then $\bm i_1,\ldots,\bm i_m,\bm j_1,\ldots,\bm j_m$ are $2m$ distinct elements chosen uniformly from $G$. Let $\bm\xi\in\{-1,1\}^G$ be a Rademacher random vector, independent of $(\Omega,\Omega')$ and $D_1,D_2,D_3$. Then we have
\begin{equation}\label{eq:pairedsample}
\sum_{\ell=1}^m\left(h\bigl((\sqrt d\,A_3\bm t)_{\bm i_\ell}\bigr)-h\bigl((\sqrt d\,A_3\bm t)_{\bm j_\ell}\bigr)\right)
\overset{\mathrm d}=
\sum_{\ell=1}^m\xi_{\bm i_\ell}\left(h\bigl((\sqrt d\,A_3\bm t)_{\bm i_\ell}\bigr)-h\bigl((\sqrt d\,A_3\bm t)_{\bm j_\ell}\bigr)\right).
\end{equation}
Applying Jensen's inequality to \eqref{eq:sampleaverage} and substituting \eqref{eq:pairedsample} yields
\begin{equation}\label{eq:clippedsigns}
 \begin{split}
 &\left\|\sup_{\bm t\in T}\Big|\frac1m\sum_{\bm i\in\Omega}h\bigl((\sqrt d\,A_3\bm t)_{\bm i}\bigr)-\frac1d\sum_{\bm i\in G}h\bigl((\sqrt d\,A_3\bm t)_{\bm{i}}\bigr)\Big|\right\|_{L^s}\\
 \le &\frac1m \left\|\sup_{\bm t\in T}\Big| \sum_{\ell=1}^m\xi_{\bm i_\ell}\left(h\bigl((\sqrt d\,A_3\bm t)_{\bm i_\ell}\bigr)-h\bigl((\sqrt d\,A_3\bm t)_{\bm j_\ell}\bigr)\right) \Big|\right\|_{L^s}\\
 \le & \frac1m\left\|\sup_{\bm t\in T}\Big|\sum_{\bm i\in\Omega}\xi_{\bm i}h\bigl((\sqrt d\,A_3\bm t)_{\bm i}\bigr)\Big|\right\|_{L^s}
 + \frac1m\left\|\sup_{\bm t\in T}\Big|\sum_{\bm j\in\Omega'}\xi_{\bm j}h\bigl((\sqrt d\,A_3\bm t)_{\bm j}\bigr)\Big|\right\|_{L^s} \\
 \le&\frac2m \left\|\sup_{\bm t\in T}\Big|\sum_{\bm i\in\Omega}\xi_{\bm i}h\bigl((\sqrt d\,A_3\bm t)_{\bm i}\bigr)\Big|\right\|_{L^s}.
 \end{split}
\end{equation}
Since $h(0)=0$ and $|h(x)-h(y)|\le2C_5\log(em)|x-y|$, applying Lemma~\ref{lem:signbound} gives
\begin{equation}\label{eq:clippedbound}
\left\|\sup_{\bm t\in T}\Big|\sum_{\bm i\in\Omega}\xi_{\bm i}h\bigl((\sqrt d\,A_3\bm t)_{\bm i}\bigr)\Big|\right\|_{L^s}\le4C_5\log(em)
\left\|\sup_{\bm t\in T}\Big|\sum_{\bm i\in\Omega}\xi_{\bm i}(\sqrt d\,A_3\bm t)_{\bm i}\Big|\right\|_{L^s}.
\end{equation}
For fixed $\Omega,\bm\xi$, $m^{-1/2}P_\Omega\bm\xi$ is a unit vector supported on $\Omega$. Applying Lemma~\ref{lem:scalarmoment} yields
\[
\begin{split}
\left\|\sup_{\bm t\in T}\Big|\sum_{\bm i\in\Omega}\xi_{\bm i}(\sqrt d\,A_3\bm t)_{\bm i}\Big|\right\|_{L^s(A_3)}&=\sqrt m\left\|\sup_{\bm t\in T}\left|\left\langle m^{-1/2}P_\Omega\bm\xi,\sqrt d\,A_3\bm t\right\rangle\right|\right\|_{L^s(A_3)} \\
&\lesssim\sqrt{m\log(em)}\bigl(w(T)+\sqrt s\bigr).
\end{split}
\]
Taking expectation with respect to $\Omega,\bm\xi$ and substituting into \eqref{eq:clippedbound} and \eqref{eq:clippedsigns} proves \eqref{eq:clippedmoment}. This completes the proof.
\end{proof}

Now we present the proof of Lemma~\ref{lem:largespace} using Lemma~\ref{lem:excesssquares} and Lemma~\ref{lem:clippedsample}.
\begin{proof}[Proof of Lemma~\ref{lem:largespace}]
Set $f(x):=(x^2-C_5^2\log^2(em))_+$ and $h(x):=\min(x^2,C_5^2\log^2(em))$. Since $x^2=h(x)+f(x)$, we have
\begin{equation}\label{eq:squaresplit}
\begin{split}
&\sup_{\bm t\in T}\bigl|\frac dm\|P_\Omega A_3\bm t\|_2^2-1\bigr|
\le\sup_{\bm t\in T}\left|\frac1m\sum_{\bm i\in\Omega}h\bigl((\sqrt d\,A_3\bm t)_{\bm i}\bigr)-\frac1d\sum_{\bm i\in G}h\bigl((\sqrt d\,A_3\bm t)_{\bm i}\bigr)\right| \\ +&\sup_{\bm t\in T}\frac1m\sum_{\bm i\in\Omega}f\bigl((\sqrt d\,A_3\bm t)_{\bm i}\bigr)+\sup_{\bm t\in T}\frac1d\sum_{\bm i\in G}f\bigl((\sqrt d\,A_3\bm t)_{\bm i}\bigr).
\end{split}
\end{equation}
Set $s:=\log(e/\delta)$. Since $\PP\{\bm i\in\Omega\}=m/d$, applying Jensen's inequality and Lemma~\ref{lem:excesssquares} yields
\begin{equation}\label{eq:totalexcess}
 \begin{split}
 &\left\|\sup_{\bm t\in T}\frac1d\sum_{\bm i\in G}f\bigl((\sqrt d\,A_3\bm t)_{\bm i}\bigr)\right\|_{L^s}
 = \left\|\sup_{\bm t\in T}\E_\Omega\frac1m\sum_{\bm i\in\Omega}f\bigl((\sqrt d\,A_3\bm t)_{\bm i}\bigr)\right\|_{L^s}\\
\le&\left\|\sup_{\bm t\in T}\frac1m\sum_{\bm i\in\Omega}f\bigl((\sqrt d\,A_3\bm t)_{\bm i}\bigr)\right\|_{L^s}
\lesssim\frac{(w(T)^2+s)\log^2(em)}m.
 \end{split}
\end{equation}
Since $m\ge C_4\eps^{-2}(w(T)^2+s)\log^3(em)$, substituting \eqref{eq:sampleexcess}, \eqref{eq:totalexcess}, and \eqref{eq:clippedmoment} into \eqref{eq:squaresplit} and choosing a sufficiently large $C_4$ gives
\[
\left\|\sup_{\bm t\in T}\bigl|\frac dm\|P_\Omega A_3\bm t\|_2^2-1\bigr|\right\|_{L^{s}}
\le\eps/e.
\]
Then applying Markov's inequality yields
\[
\PP\left\{\sup_{\bm t\in T}\bigl|\frac dm\|P_\Omega A_3\bm t\|_2^2-1\bigr|>\eps\right\}
\le e^{-\log(e/\delta)}=\delta/e<\delta.
\]
This completes the proof.
\end{proof}

\subsection{Proof of Lemma~\ref{lem:largeentries}}

Throughout this subsection, set $k:=\lceil m+\log d\rceil$ as in Lemma~\ref{lem:largeentries}.
We first establish a Bernoulli comparison result in the following lemma.

\begin{lemma}\label{lem:normalsum}
Let $0\preceq\Sigma\preceq3I_d$, and let
$\bm z_0\sim N(\bm{0},\Sigma)$. For $a>0$, define
\[
 I_0:=\{i\in[d]:|{z}_{0,i}|>a\},\qquad
 p:=\PP\{|N(0,3)|>a/2\}.
\]
Let $R_p\in\R^{d\times d}$ be a diagonal matrix with independent Bernoulli-$p$
entries. If $p\le1/6$, then every deterministic real symmetric
matrix $L\in\R^{d\times d}$ satisfies
\begin{equation}\label{eq:normalsum}
\|P_{I_0}LP_{I_0}\|_{2,L^k}
\le2\|R_pLR_p\|_{2,L^k}.
\end{equation}
\end{lemma}

\begin{proof}
Suppose $\bm z_1\sim N(\bm{0},3I_d-\Sigma)$ is a random vector independent of $\bm z_0$. Then ${z}_{0,i}+{z}_{1,i}\sim N(0,3)$.
Thus, set $I_1:=\{i\in[d]:|{z}_{0,i}+{z}_{1,i}|>a/2\}$, and then we have $P_{I_1}\overset{\mathrm d}=R_p$.
Fix $\bm z_0$. If $i\in I_0\setminus I_1$, then
$|{z}_{0,i}|>a$ and $|{z}_{0,i}+{z}_{1,i}|\le a/2$, and hence $|{z}_{1,i}|>a/2$. Since $\operatorname{Var}({z}_{1,i})\le3$, by the definition of $p$, we obtain
\begin{equation}\label{eq:missedentries}
\E_{\bm z_1} P_{I_0\setminus I_1}
\preceq pP_{I_0}.
\end{equation}
Since $P_{I_0}P_{I_1}=P_{I_0}-P_{I_0\setminus I_1}$, we have
\begin{equation}\label{eq:matrixsplit}
P_{I_0}LP_{I_0}-P_{I_0}P_{I_1}LP_{I_1}P_{I_0}
=P_{I_0\setminus I_1}LP_{I_0}+P_{I_0}LP_{I_0\setminus I_1}-P_{I_0\setminus I_1}LP_{I_0\setminus I_1}.
\end{equation}
For $P_{I_0\setminus I_1}LP_{I_0}$, \eqref{eq:missedentries} yields
\[
\left\|\E_{\bm z_1}P_{I_0\setminus I_1}LP_{I_0}\right\|_2
=\left\|\E_{\bm z_1}P_{I_0\setminus I_1}P_{I_0}LP_{I_0}\right\|_2
\le p\|P_{I_0}LP_{I_0}\|_2.
\]
Similarly, $P_{I_0}LP_{I_0\setminus I_1}$ is its transpose and obeys the same estimate.
Since $-\|P_{I_0}LP_{I_0}\|_2P_{I_0\setminus I_1} \preceq P_{I_0\setminus I_1}LP_{I_0\setminus I_1} \preceq\|P_{I_0}LP_{I_0}\|_2P_{I_0\setminus I_1}$,
taking conditional expectations gives
\begin{equation}\label{eq:thirdterm}
\left\|\E_{\bm z_1}P_{I_0\setminus I_1}LP_{I_0\setminus I_1}
\right\|_2
\le p\|P_{I_0}LP_{I_0}\|_2.
\end{equation}
Hence, \eqref{eq:matrixsplit} gives
\[
\begin{split}
\|P_{I_0}LP_{I_0}\|_2
&\le\left\|\E_{\bm z_1}P_{I_0}P_{I_1}LP_{I_1}P_{I_0}\right\|_2+3p\|P_{I_0}LP_{I_0}\|_2\\
&\le\|P_{I_0}P_{I_1}LP_{I_1}P_{I_0}\|_{2,L^1(\bm z_1)}+3p\|P_{I_0}LP_{I_0}\|_2 \le\|P_{I_1}LP_{I_1}\|_{2,L^1(\bm z_1)}+3p\|P_{I_0}LP_{I_0}\|_2,
\end{split}
\]
where the second inequality follows from Jensen's inequality. Since $p\le1/6$, Jensen's inequality yields
\[
\|P_{I_0}LP_{I_0}\|_{2,L^k}
\le2\Bigl\|\|P_{I_1}LP_{I_1}\|_{2,L^1(\bm z_1)}\Bigr\|_{L^k} \le2\|P_{I_1}LP_{I_1}\|_{2,L^k}
=2\|R_pLR_p\|_{2,L^k}.
\]
\end{proof}

The following lemma contains several needed estimates concerning random diagonal projections.

\begin{lemma}[{\cite[Propositions~6--9]{TroppPaving}}]\label{lem:matrixsample}
Suppose $p\ge 2\log d$ is an even integer. For a matrix $X$, denote $\|X\|_{\max}:=\max_{i,j}|X_{ij}|$. Let $0<\theta<1$, and let $R_\theta,R'_\theta\in\R^{d\times d}$ be independent diagonal matrices with independent Bernoulli-$\theta$ entries. Then, for every $X\in\R^{d\times d}$,
\begin{equation}\label{eq:columnsample}
\|XR_\theta\|_{2,L^{p}}
\lesssim\sqrt p\Bigl\|\max_{1\le j\le d}\|XR_\theta\bm e_j\|_2\Bigr\|_{L^{p}}+\sqrt\theta\|X\|_2,
\end{equation}
\begin{equation}\label{eq:columnnorms}
\Bigl\|\max_{1\le j\le d}\|R_\theta X\bm e_j\|_2\Bigr\|_{L^{p}}
\lesssim\sqrt p\|X\|_{\max}+\sqrt\theta\max_{1\le j\le d}\|X\bm e_j\|_2.
\end{equation}
Moreover, for every $0<\rho<1/2$, we have
\begin{equation}\label{eq:densitychange}
\|R_\theta XR_\theta\|_{2,L^{p}}
\lesssim\theta^{1/4}\|X\|_2+\rho^{-1/4}\|R_\rho XR_\rho\|_{2,L^{p}}.
\end{equation}
If every diagonal entry of $X$ is zero, then also
\begin{equation}\label{eq:matrixpair}
\|R_\theta XR_\theta\|_{2,L^{p}}
\lesssim\|R_\theta XR'_\theta\|_{2,L^{p}}.
\end{equation}
\end{lemma}

In the following lemma, we combine Lemma~\ref{lem:normalsum} and Lemma~\ref{lem:matrixsample} to control the moment of the form in Lemma~\ref{lem:largeentries} when the off-diagonal entries are sufficiently small.

\begin{lemma}\label{lem:fixedmatrix}
Let $\Sigma\in\R^{d\times d}$ with $\Sigma\succeq0$, $\|\Sigma\|_2\le m$, and $\max_i\Sigma_{ii}\le2$. Suppose that
\begin{equation}\label{eq:smallentries}
\max_{i\ne j}|\Sigma_{ij}|\le c(m+2)^{-17}k^{-3/2}.
\end{equation}
Set $\bm z\sim N(\bm{0},\Sigma)$ and $I:=\{i:|{z}_i|>C_2\sqrt{\log(em)}\}$. Then
\begin{equation}\label{eq:fixedmoment}
\|P_I\Sigma P_I\|_{2,L^k}\le3.
\end{equation}
Here $c$ is a universal constant and $C_2$ is the universal constant in Lemma~\ref{lem:largeentries}.
\end{lemma}

\begin{proof}
Recall that $k:=\lceil m+\log d\rceil$. If $m\le 3$, then \eqref{eq:fixedmoment} follows from $\|\Sigma\|_2\le m$. Thus assume $m>3$. Denote the off-diagonal part of $\Sigma$ by $\Sigma^0:=\Sigma-\diag\Sigma$. Then $\Sigma^0$ is symmetric, with zero diagonal entries, and $\|\Sigma^0\|_2\le m+2$. If $\Sigma$ is diagonal, the conclusion follows from $\|P_I\Sigma P_I\|_2\le2$. Assume $\Sigma^0\neq 0$. Set $\rho:=k\|\Sigma^0\|_{\max}^2/(m+2)^2$.
Choosing $c\le1$, the hypothesis in \eqref{eq:smallentries} gives $0<\rho<1/2$ and $\rho^{-1/4}k\|\Sigma^0\|_{\max}\le\sqrt{c}(m+2)^{-8}$.

We claim that
\begin{equation}\label{eq:sparsematrix}
\|R_\rho\Sigma^0 R_\rho\|_{2,L^{4k}}
\lesssim k\|\Sigma^0\|_{\max}.
\end{equation}
Fix a sufficiently large absolute constant $C_2$, and set
$t:=(C_2(m+2))^{-4}$. Let $R_t$ be a Bernoulli random projection
with parameter $t$, independent of $\bm z$.
Applying \eqref{eq:densitychange} to $R_t$ and then using
\eqref{eq:sparsematrix} gives
\[
\|R_t\Sigma^0 R_t\|_{2,L^{4k}}
\lesssim(m+2)t^{1/4}+\rho^{-1/4}k\|\Sigma^0\|_{\max}
\le C_2^{-1}+\sqrt{c}(m+2)^{-8}.
\]
For $C_2$ sufficiently large and $c$ sufficiently small,
this implies
\[
\|R_t\Sigma^0 R_t\|_{2,L^{4k}}
\le C_2^{-3/4}.
\]
Since $k\ge4$ and $m+2\le k\le e^{k/2}$, Markov's inequality gives
\begin{equation}\label{eq:matrixtail}
\PP_{R_t}\{\|R_t\Sigma^0 R_t\|_2>1\}
\le\|R_t\Sigma^0 R_t\|_{2,L^{4k}}^{4k}
\le C_2^{-3k}\le t^2/2,
\end{equation}
where the last inequality follows from $t=(C_2(m+2))^{-4}$ and $k:=\lceil m+\log d\rceil$.

Set $a:=C_2\sqrt{\log(em)}$ and $p:=\PP\{|N(0,3)|>a/2\}$. Then the Gaussian tail bound gives
$p\le2e^{-a^2/24}\le2(em)^{-C_2^2/24}$.
Applying \eqref{eq:densitychange} to $R_p$ and using
\eqref{eq:sparsematrix} gives
\[
\|R_p\Sigma^0 R_p\|_{2,L^{4k}}
\lesssim(m+2)p^{1/4}+\sqrt{c}(m+2)^{-8}\le t^2/4,
\]
where the last inequality follows by choosing $C_2$ sufficiently large and
$c$ sufficiently small.

Conditional on $R_t$, the covariance of $R_t\bm z$ is $R_t\Sigma R_t$. For each fixed $R_t$ satisfying $\|R_t\Sigma^0 R_t\|_2\le1$, we have $\|R_t\Sigma R_t\|_2 \le 2+1=3$. The projection onto the coordinates where
$|(R_t\bm z)_i|>a$ is $R_tP_I$.
Applying Lemma~\ref{lem:normalsum} to $R_t\bm z$
and $R_t\Sigma^0 R_t$ gives
\begin{equation}\label{eq:localbound}
\|R_tP_I\Sigma^0 P_I R_t\|_{2,L^k(\bm z)}
\le2\|R_tR_p\Sigma^0 R_pR_t\|_{2,L^k(R_p)}
\le2\|R_p\Sigma^0 R_p\|_{2,L^{4k}(R_p)}
\le t^2/2.
\end{equation}

For each $\bm z$, the diagonal entries of $P_I\Sigma^0 P_I$ are zero. Thus by the definition of $R_t$,
\[
\E_{R_t}R_tP_I\Sigma^0 P_I R_t=t^2P_I\Sigma^0 P_I.
\]
Splitting the expectation over $\{\|R_t\Sigma^0 R_t\|_2\le1\}$ and $\{\|R_t\Sigma^0 R_t\|_2>1\}$ yields
\begin{equation}\label{eq:scalebound}
\begin{split}
t^2\|P_I\Sigma^0 P_I\|_2 &\le\bigl\|1_{\{\|R_t\Sigma^0 R_t\|_2\le1\}}R_tP_I\Sigma^0 P_I R_t\bigr\|_{2,L^1(R_t)}+\PP_{R_t}\{\|R_t\Sigma^0 R_t\|_2>1\}\|P_I\Sigma^0 P_I\|_2\\
&\le\bigl\|1_{\{\|R_t\Sigma^0 R_t\|_2\le1\}}R_tP_I\Sigma^0 P_I R_t\bigr\|_{2,L^1(R_t)}+\frac{t^2}{2}\|P_I\Sigma^0 P_I\|_2,
\end{split}
\end{equation}
where the last inequality follows from \eqref{eq:matrixtail}. Since $R_t$ is independent of $\bm z$, \eqref{eq:scalebound} and Minkowski's inequality give
\begin{equation}\label{eq:fullbound}
\begin{split}
\|P_I\Sigma^0 P_I\|_{2,L^k}
&\le\frac2{t^2}\Bigl\|\bigl\|1_{\{\|R_t\Sigma^0 R_t\|_2\le1\}}R_tP_I\Sigma^0 P_I R_t\bigr\|_{2,L^1(R_t)}\Bigr\|_{L^k}\\
&\le\frac2{t^2}\Bigl\|1_{\{\|R_t\Sigma^0 R_t\|_2\le1\}}\|R_tP_I\Sigma^0 P_I R_t\|_{2,L^k(\bm z)}\Bigr\|_{L^1}\le\frac2{t^2}\cdot\frac{t^2}{2}=1,
\end{split}
\end{equation}
where the last inequality follows from \eqref{eq:localbound}. Since $\|P_I(\diag\Sigma)P_I\|_2\le2$, we have
\[
\|P_I\Sigma P_I\|_{2,L^k}
\le\|P_I\Sigma^0 P_I\|_{2,L^k}+2\le1+2=3.
\]

It remains to prove the claim in \eqref{eq:sparsematrix}.
Let $R'_\rho$ be an independent copy of $R_\rho$.
Applying Lemma~\ref{lem:matrixsample} gives
\begin{equation}\label{eq:pairmoment}
\begin{split}
\|R_\rho\Sigma^0 R_\rho\|_{2,L^{4k}}
&\lesssim\|R_\rho\Sigma^0 R'_\rho\|_{2,L^{4k}}\\
&\lesssim\Bigl\|\sqrt k\bigl\|\max_j\|R_\rho\Sigma^0 R'_\rho\bm e_j\|_2\bigr\|_{L^{4k}(R'_\rho)}+\sqrt\rho\|R_\rho\Sigma^0\|_2\Bigr\|_{L^{4k}}\\
&\le\sqrt k\bigl\|\max_j\|R_\rho\Sigma^0 R'_\rho\bm e_j\|_2\bigr\|_{L^{4k}}+\sqrt\rho\|R_\rho\Sigma^0\|_{2,L^{4k}}\\
&\le\sqrt k\bigl\|\max_j\|R_\rho\Sigma^0\bm e_j\|_2\bigr\|_{L^{4k}}+\sqrt\rho\|\Sigma^0 R_\rho\|_{2,L^{4k}}.
\end{split}
\end{equation}
Since $\|\Sigma^0\|_2\le m+2$, Lemma~\ref{lem:matrixsample} yields
\begin{equation}\label{eq:columnmoment}
\bigl\|\max_j\|R_\rho\Sigma^0\bm e_j\|_2\bigr\|_{L^{4k}}
\lesssim\sqrt k\|\Sigma^0\|_{\max}+\sqrt\rho(m+2),
\end{equation}
\begin{equation}\label{eq:samplemoment}
\|\Sigma^0 R_\rho\|_{2,L^{4k}}
\lesssim\sqrt k\|\Sigma^0\|_2 +\sqrt\rho\|\Sigma^0\|_2 \le(\sqrt k+\sqrt\rho)(m+2).
\end{equation}
Substituting \eqref{eq:columnmoment} and
\eqref{eq:samplemoment} into
\eqref{eq:pairmoment} gives
\[
\|R_\rho\Sigma^0 R_\rho\|_{2,L^{4k}}
\lesssim k\|\Sigma^0\|_{\max}+(m+2)\sqrt{\rho k}=2k\|\Sigma^0\|_{\max},
\]
where the last equality follows from the definition $\rho=k\|\Sigma^0\|_{\max}^2/(m+2)^2$.
This completes the proof.
\end{proof}

The following lemma is a concentration inequality that we use to bound the probability that the hypotheses in Lemma~\ref{lem:fixedmatrix} hold.

\begin{lemma}[{\cite[Theorem~1.1]{HW}}]\label{lem:quadratictail}
Let $\bm\xi\in\{-1,1\}^n$ be a random vector with independent Rademacher coordinates. Then, for every deterministic matrix $M\in\R^{n\times n}$ and every $u>0$,
\begin{equation}\label{eq:quadratictail}
\PP\left\{\left|\bm\xi^{\mathsf T}M\bm\xi-\tr M\right|>C\left(\sqrt u\,\|M\|_{\mathrm F}+u\|M\|_2\right)\right\}
\le2e^{-u},
\end{equation}
where $C$ is a universal constant.
\end{lemma}

Now we present the proof of Lemma~\ref{lem:largeentries} based on Lemma~\ref{lem:fixedmatrix} and Lemma~\ref{lem:quadratictail}.
\begin{proof}[Proof of Lemma~\ref{lem:largeentries}]

Write $D=\diag(\bm\xi)$. For $\bm i,\bm j\in G$, define $M_{\bm i,\bm j}\in \R^{G\times G}$ by
\[
(M_{\bm i,\bm j})_{\bm x,\bm y}
:=d^{-1}(-1)^{\bm i\cdot\bm x+\bm j\cdot\bm y}(\Sigma_0)_{\bm x,\bm y}.
\]
Since $\Sigma_1=HD\Sigma_0DH$ and $(\Sigma_0)_{\bm x,\bm x}=1$, we have
\[
\begin{split}
(\Sigma_1)_{\bm i,\bm j}
&=d^{-1}\sum_{\bm x,\bm y\in G}
(-1)^{\bm i\cdot\bm x+\bm j\cdot\bm y}
{\xi}_{\bm x}(\Sigma_0)_{\bm x,\bm y}{\xi}_{\bm y}
=\bm\xi^{\mathsf T}M_{\bm i,\bm j}\bm\xi,\\
\tr M_{\bm i,\bm j}
&=d^{-1}\sum_{\bm x\in G}(-1)^{(\bm i+\bm j)\cdot\bm x}
=1_{\{\bm i=\bm j\}}.
\end{split}
\]
Since $\sum_{\bm j}|(\Sigma_0)_{\bm i,\bm j}|\le m$ for every $\bm i\in G$, Gershgorin's circle theorem
gives $\|\Sigma_0\|_2\le m$.
Moreover, $\|\Sigma_0\|_{\mathrm F}^2\le \|\Sigma_0\|_2\tr\Sigma_0\le m d$. Since multiplication by diagonal sign matrices preserves the two matrix
norms, we have $\|M_{\bm i,\bm j}\|_2=d^{-1}\|\Sigma_0\|_2\le m/d$
and $\|M_{\bm i,\bm j}\|_{\mathrm F}^2
=d^{-2}\|\Sigma_0\|_{\mathrm F}^2\le m/d$.
Choose the tail parameter $u:=k\log(m)+2\log(d)$ and define the event $\mathcal E$ as
\begin{equation}\label{eq:entrybound}
\mathcal E:=\left\{\max_{\bm i,\bm j\in G}|(\Sigma_1-I_d)_{\bm i,\bm j}|
 \le C_5\Big(\sqrt{\frac{m u}{d}}+\frac{m u}{d}\Big)\right\}.
\end{equation}
Applying Lemma~\ref{lem:quadratictail} to every $M_{\bm i,\bm j}$ and then taking a union bound gives
\begin{equation}\label{eq:failurebound}
 \PP_D(\mathcal E^c)
 \le2d^2e^{-u}=2m^{-k}.
\end{equation}

Recall that $d\ge(C_3m)^{40}$. If $\Sigma_1$ satisfies \eqref{eq:entrybound}, then it satisfies all hypotheses in Lemma~\ref{lem:fixedmatrix}.
Applying Lemma~\ref{lem:fixedmatrix} gives $\|P_I\Sigma_1P_I\|_{2,L^k(\bm z)}\le3$ if $\Sigma_1$ satisfies \eqref{eq:entrybound}.
Since $\|P_I\Sigma_1P_I\|_2\le\|\Sigma_1\|_2\le m$ for every $\Sigma_1$, we have
\begin{equation}\label{eq:matrixmoment}
 \|P_I\Sigma_1P_I\|_{2,L^k}^k
 \le3^k\PP_D(\mathcal E)+m^k\PP_D(\mathcal E^c)
 \le3^k+2\le4^k.
\end{equation}
This completes the proof.
\end{proof}

\section{Proof of Theorem~\ref{thm:twoblocks}}\label{sec:blockproof}

We prove Theorem~\ref{thm:twoblocks} by constructing a finite family of perturbations near two opposite unit vectors.
The following lemma bounds the probability of an event used in the proof of Theorem~\ref{thm:twoblocks}.

\begin{lemma}\label{lem:largecount}
Suppose $k$ is a positive integer. Let $d=2^r$, and let
$H\in\R^{d\times d}$ be orthogonal with entries in
$\{-d^{-1/2},d^{-1/2}\}$. Let $\bm\xi\in\{-1,1\}^d$ have
independent Rademacher coordinates. Set $\bm y:=H\bm\xi$.
Then
\begin{equation}\label{eq:countbound}
\lim_{r\to\infty}\PP\left\{\left|\left\{i\in[d]:  y_i>\frac14\sqrt{\log d}\right\}\right|\ge k\right\}=1,
\end{equation}
uniformly over all such $H$.
\end{lemma}

\begin{proof}
Set $\lambda:=\tfrac23\sqrt{\log d}$. Define $S_1:=\sum_{i=1}^d e^{\lambda y_i}$ and $S_2:=\sum_{i=1}^d e^{2\lambda y_i}$.
We first establish the moment estimates
\begin{equation}\label{eq:countmoments}
 \E S_1= d^{11/9}(1+o(1)),\qquad
 \operatorname{Var}(S_1)\le\E S_2\le d^{17/9}.
\end{equation}

The independence of the coordinates of $\bm\xi$ yields
\[
\E S_1 =\sum_{i=1}^d\prod_{j=1}^d\E e^{\lambda H_{ij}\xi_j}
 =\sum_{i=1}^d\prod_{j=1}^d\frac{e^{\lambda H_{ij}}+e^{-\lambda H_{ij}}}{2}
 =d\left(\frac{e^{\lambda/\sqrt d}+e^{-\lambda/\sqrt d}}{2}\right)^d.
\]
Using Taylor expansion gives $\log((e^u+e^{-u})/2)=u^2/2+O(u^4)$. Since $\lambda/\sqrt d\to0$, we have
\[
\log\frac{\E S_1}{d}
=\frac{\lambda^2}{2} +O\left(\frac{\lambda^4}{d}\right)
=\frac29\log d+O\left(\frac{(\log d)^2}{d}\right),
\]
which proves the first estimate in \eqref{eq:countmoments}.

We next estimate the variance. For $i\ne j$, the orthogonality of the rows of $H$ implies that rows $i$ and $j$ have equal entries in exactly $d/2$ coordinates and opposite entries in exactly $d/2$ coordinates. Therefore we have
\begin{equation}\label{eq:pairbound}
\E e^{\lambda( y_i+ y_j)}=\left(\frac{e^{2\lambda/\sqrt d}+e^{-2\lambda/\sqrt d}}{2}\right)^{d/2}\le\left(\frac{e^{\lambda/\sqrt d}+e^{-\lambda/\sqrt d}}{2}\right)^{2d}
=\E e^{\lambda y_i}\E e^{\lambda y_j}.
\end{equation}
Then \eqref{eq:pairbound} gives $\operatorname{Cov}( e^{\lambda y_i},e^{\lambda y_j})\le0$. Therefore, we obtain
\[
\operatorname{Var}(S_1)\le\sum_{i=1}^d\operatorname{Var}( e^{\lambda y_i})
\le\sum_{i=1}^d\E e^{2\lambda y_i}=\E S_2.
\]
Moreover, since $(e^u+e^{-u})/2\le e^{u^2/2}$ for $u\in\R$, we have
\[
\E S_2 =d\left(\frac{e^{2\lambda/\sqrt d}+e^{-2\lambda/\sqrt d}}{2}\right)^d
\le d e^{2\lambda^2}=d^{17/9}.
\]

Set $n:=|\{i\in[d]: y_i>\tfrac14\sqrt{\log d}\}|$. If a coordinate $ y_i$ satisfies $ y_i\le\tfrac14\sqrt{\log d}$, then $e^{\lambda y_i}\le d^{1/6}$. Applying the Schwarz inequality to the remaining coordinates gives
\begin{equation}\label{eq:countsplit}
 S_1\le d^{7/6}+\sqrt{nS_2}.
\end{equation}
Note that \eqref{eq:countmoments} implies that $d^{7/6}\le\E S_1/4$ for all sufficiently large $r$. If $n<k$ and $S_1\ge\E S_1/2$, then \eqref{eq:countsplit} gives $\E S_1/4\le\sqrt{nS_2}<\sqrt{kS_2}$. Applying Chebyshev's inequality and Markov's inequality yields
\[
 \PP\{n<k\}\le\PP\{S_1<\frac{\E S_1}{2}\}+\PP\left\{\sqrt{kS_2}>\frac{\E S_1}{4}\right\}
 \le\frac{4\operatorname{Var}(S_1)}{(\E S_1)^2}+\frac{16k\E S_2}{(\E S_1)^2}
 \lesssim (4+16k)d^{-5/9}.
\]
This completes the proof.
\end{proof}

\begin{proof}[Proof of Theorem~\ref{thm:twoblocks}]
Recall that $d=2^r$ and $G=\F_2^r$. Suppose $r$ is sufficiently large such that $r>2^m$ and $m\log d>64$.

For an $m$-dimensional subspace $V\subset G$ and $\bm c,\bm c'\in G$, the sets $\bm c+V$ and $\bm c'+V$ are either equal or disjoint.
Define
\[
E_V:=\{(\bm c_1+V)\cup\cdots\cup (\bm c_m+V): \bm c_1+V,\ldots,\bm c_m+V \text{ are }m\text{ distinct sets different from }V\},
\]
\[
U:=\left\{\frac{1}{\sqrt{m2^m}}\sum_{\bm x\in S}\pm\bm e_{\bm x}:V\subset G\text{ is an }m\text{-dimensional subspace, }S\in E_V\right\},
\]
where $\bm e_{\bm x}\in\R^G$ is the coordinate vector at $\bm x$. Then $U$ is a set of unit vectors. Since $\bm0\notin S$ for every $V$ and every $S\in E_V$, $\langle\bm u,\bm{e}_{\bm0}\rangle=0$ for every $\bm{u}\in U$. Since all choices of the signs before $\bm e_{\bm x}$ occur in the definition of $U$, we have $D_1U=U$ for every $D_1$, and in particular $U=-U$.

Set $\theta:=8/\sqrt{m\log d}$ and
$a:=\sqrt{1-\theta^2}$. Define
\begin{equation}\label{eq:smallset}
T_{m,r}:=\left\{\pm(a\bm e_{\bm0}+\theta\bm u):\bm u\in U\right\}.
\end{equation}
Then $T_{m,r}$ is a finite symmetric subset of $\mathbb{S}^{d-1}$. Since $D_1U=U$, $D_1T_{m,r}=T_{m,r}$.

We first verify the width bound of $T_{m,r}$. There are fewer than $d^m$ $m$-dimensional subspaces $V\subset G$. For each $V$, there are fewer than $d^m$ elements in $E_V$, and there are $2^{m2^m}$ sign choices for each element in $E_V$. Hence $|U|\le d^{2m}2^{m2^m}$ and $\log|U|\le2m\log d+m2^m\log2\le3m\log d$. Using the standard bounds on the Gaussian width of finite sets $w(U)\le\sqrt{2\log|U|}$, we have
\begin{equation}\label{eq:setwidth}
 w(T_{m,r}) \le aw(\{\pm \bm e_{\bm0}\})+ \theta w(U)
 \le\sqrt{2\log 2}+\theta\sqrt{2\log|U|}
 \le\sqrt{2\log 2}+8\sqrt6<21.
\end{equation}

Fix an arbitrary $m$-element set $\Omega\subset G$, and choose an $m$-dimensional subspace $V\subset G$ containing $\Omega$. Write $G/V:=\{\bm c+V:\bm c\in G\}$. Then $|G/V|=|V^\perp|=2^{r-m}$. Set $A:=\sqrt{d/m}\,P_\Omega HD_2H$.
For each $C=\bm c+V\in G/V$, define
\[
 y_C:=2^{(m-r)/2}\sum_{\bm b\in V^\perp}\xi^{(2)}_{\bm b}(-1)^{\bm c\cdot\bm b},
\]
where $(\bm1_C)_{\bm x}=1_{\{\bm x\in C\}}$ and
$D_2=\diag(\bm\xi^{(2)})$.
The value of $ y_C$ does not depend on the representative:
if $\bm c'=\bm c+\bm v$ with $\bm v\in V$, then
$\bm c'\cdot\bm b=\bm c\cdot\bm b$ for every
$\bm b\in V^\perp$.
Define the event
\begin{equation}\label{eq:selectevent}
\mathcal E_\Omega:=
\left\{\bigg|\Big\{C\in G/V: y_C>\tfrac14\sqrt{\log |V^\perp|}\Big\}\bigg|\ge m+1\right\}.
\end{equation}
On $\mathcal E_\Omega$, choose distinct $C_1,\ldots,C_m\ne V$ with $ y_{C_j}>\sqrt{\log |V^\perp|}/4$, and set
\[
\bm u_\Omega:=\frac1{\sqrt{m2^m}}\sum_{i=1}^m\bm1_{C_i}.
\]
Then $\bm u_\Omega\in U$ by the definition of $U$. Choose $\eps_{\Omega}\in\{-1,1\}$ such that $\eps_{\Omega}\langle A\bm e_{\bm0},A\bm u_\Omega\rangle\ge0$, and define $\bm t_\Omega:=D_1(a\bm e_{\bm0}+\theta\eps_\Omega\bm u_\Omega)\in T_{m,r}$.
We claim that on the event $\mathcal E_\Omega$,
\begin{equation}\label{eq:largeimage}
 \|AD_1\bm t_\Omega\|_2^2>2.
\end{equation}

We next bound the probability of the event $\mathcal E_\Omega$. Define the coefficient matrix $H_V$ with $2^{r-m}$ rows and columns by $(H_V)_{\bm c,\bm b}:=2^{(m-r)/2}(-1)^{\bm c\cdot\bm b}$, where $\bm c+V \in G/V$ and $\bm b\in V^\perp$. Since $V^\perp$ is a subspace of $G$ and $(V^\perp)^\perp=V$, we have
\[
2^{m-r}\sum_{\bm b\in V^\perp}(-1)^{(\bm c+\bm c')\cdot\bm b}
=1_{\{\bm c+\bm c'\in V\}}.
\]
Thus $H_V$ is an orthogonal matrix. Note that the probability bound in Lemma~\ref{lem:largecount} is uniform with respect to the orthogonal matrix. Applying Lemma~\ref{lem:largecount} to $H_V$ yields
\begin{equation}\label{eq:eventlimit}
\lim_{r\to\infty}
\sup_{\Omega\subset G, |\Omega|=m}\PP(\mathcal E_\Omega^c)=0.
\end{equation}
Together with the claim, this proves Theorem~\ref{thm:twoblocks}.

It remains to prove the claim. Since $V$ is a subspace, for $C=\bm{c}+V$,
\[
(H\bm 1_{C})_{\bm b}=2^{-r/2}\sum_{\bm v\in V}(-1)^{(\bm c+\bm v)\cdot\bm b}
 =2^{-r/2}(-1)^{\bm c\cdot\bm b}\sum_{\bm v\in V}(-1)^{\bm v\cdot\bm b}
 =2^{m-r/2}(-1)^{\bm c\cdot\bm b}1_{\{\bm b\in V^\perp\}}.
\]
For $\bm x\in\Omega\subset V$ and $\bm b\in V^\perp$, $\bm x\cdot\bm b=0$. Thus for $\bm x\in\Omega$, we have
\begin{equation}\label{eq:samplevectors}
(2^{-m/2}A\bm 1_{C})_{\bm x}=\sqrt{\frac dm}\sum_{\bm b\in V^\perp}\frac{(-1)^{\bm x\cdot\bm b}}{\sqrt d}\xi^{(2)}_{\bm b}2^{(m-r)/2}(-1)^{\bm c\cdot\bm b}
=\frac{ y_C}{\sqrt m}.
\end{equation}
In particular, all sampled coordinates of $2^{-m/2}A\bm 1_{C}$ are equal. On $\mathcal E_\Omega$, the definition of $\bm u_\Omega$ gives $(A\bm u_\Omega)_{\bm x}=m^{-1}\sum_{j=1}^m y_{C_j}$ for every $\bm x\in\Omega$. Hence,
\begin{equation}\label{eq:imagesize}
 \|A\bm u_\Omega\|_2^2
 =\frac1m\Big(\sum_{j=1}^m y_{C_j}\Big)^2
 >\frac1m\left(\frac m4\sqrt{\log(2^{r-m})}\right)^2
 =\frac{m\log(2^{r-m})}{16}.
\end{equation}
Since $D_1^2=I$, we have $AD_1\bm t_\Omega=A(a\bm e_{\bm0}+\theta\eps_\Omega\bm u_\Omega)$. By the choice of $\eps_\Omega$,
\[
\begin{split}
\|AD_1\bm t_\Omega\|_2^2&=a^2\|A\bm e_{\bm0}\|_2^2+\theta^2\|A\bm u_\Omega\|_2^2+2a\theta\eps_\Omega\langle A\bm e_{\bm0},A\bm u_\Omega\rangle\\
&\ge\theta^2\|A\bm u_\Omega\|_2^2>\frac{64}{m\log d}\,\frac{m\log(2^{r-m})}{16}=4\left(1-\frac mr\right)>2.
\end{split}
\]
This proves the claim, and the proof is completed.
\end{proof}

\bibliographystyle{abbrv}
\bibliography{hreference}

\end{document}